\documentclass[preprint,preprintnumbers]{revtex4-1}
\usepackage{amssymb, amsthm}
\usepackage{amsmath}    
\usepackage{mathrsfs}   
\usepackage{graphicx}   
\usepackage{color}      
\usepackage{footnote}   

\usepackage{slashed}    
\usepackage{subfigure}  
\usepackage{hyperref}   

\usepackage{rotating}   
\usepackage{multirow}   

\begin{document}
 
\preprint{NT@UW-16-07}
\title{Validity of the Weizs\"{a}cker-Williams approximation and the analysis of beam dump experiments: Production of a new scalar boson}
\author{Yu-Sheng Liu}\email{mestelqure@gmail.com} 
\author{David McKeen}\email{dmckeen@pitt.edu}
\author{Gerald A. Miller}\email{miller@phys.washington.edu}
\affiliation{Department of Physics,	University of Washington, Seattle, Washington 98195-1560, U.S.A.}

\begin{abstract}
Beam dump experiments have been used to search for new particles with null results interpreted in terms of limits on masses $m_\phi$ and coupling constants $\epsilon$. However these limits have been obtained by using approximations [including the Weizs\"{a}cker-Williams (WW) approximation] or Monte-Carlo simulations. We display methods, using a new scalar boson as an example, to obtain the cross section and the resulting particle production numbers without using approximations or Monte-Carlo simulations. We show that the approximations cannot be used to obtain accurate values of cross sections. The corresponding exclusion plots differ by substantial amounts when seen on a linear scale. In the event of a discovery, we generate pseudodata (assuming given values of $m_\phi$ and $\epsilon$) in the currently allowed regions of parameter space. The use of approximations to analyze the pseudodata for the future experiments is shown to lead to considerable errors in determining the parameters.  Furthermore, a new region of parameter space can be explored without using one of the common approximations, $m_\phi\gg m_e$. Our method can be used as a consistency check for Monte-Carlo simulations.
\end{abstract}
\maketitle


\section{introduction}

Beam dump experiments have been aimed at searching for new particles, such as dark photons and axions (see, e.g. \cite{Essig:2013lka} and references therein) that decay to lepton and/or photon pairs. Electron beam dumps in particular have received a large amount of theoretical attention in recent years~\cite{Bjorken:2009mm,Andreas:2012mt}. The typical setup of an electron beam dump experiment is to dump an electron beam into a target, in which the electrons are stopped (For a discussion of proton beam dumps, which is beyond the scope of this work, see, e.g.~\cite{Blumlein:2013cua,deNiverville:2016rqh}). The new particles produced by the bremsstrahlung-like process pass through a shield region and decay. These new particles can be detected by their decay products, electron and/or photon pairs, measured by the detector downstream of the decay region. Previous work simplified the necessary phase space integral by using the Weizs\"{a}cker-Williams (WW) approximation \cite{vonWeizsacker:1934nji,Williams:1935dka} which, also known as method of virtual quanta, is a semiclassical approximation. The idea is that the electromagnetic field generated by a fast moving charged particle is nearly transverse which is like a plane wave and can be approximated by real photon. The use of the  WW approximation in bremsstrahlung processes was developed in Refs.~\cite{Kim:1973he,Tsai:1973py} and applied to beam dump experiments in Refs.~\cite{Bjorken:2009mm,Tsai:1986tx}. The WW approximation simplifies evaluation of the integral over phase space and approximates the 2 particle to 3 particle (2 to 3) cross section in terms of a 2 particle to 2 particle (2 to 2) cross section. For the WW approximation to work in a beam dump experiment, it needs the incoming beam energy to be much greater than the mass of the new particle, $m_\phi$, and electron mass $m_e$. 

The previous work \cite{Bjorken:2009mm} used the following three approximations:
\begin{enumerate}
\item WW approximation;
\item a further simplification of the phase space integral, see Eq. (\ref{eq:tmin tmax});
\item $m_\phi\gg m_e$.
\end{enumerate}
The combination of the first two approximations has been denoted \cite{Kim:1973he} the improved WW (IWW) approximation. The name ``improved WW" might be somewhat misleading since the procedure reduces the computational time but not to improve accuracy). In this paper, we will focus on examining the validity of WW and IWW approximations (The validity of WW approximation is also discussed in other processes, e.g.~\cite{Brodsky:1971ud}). The third approximation used to simplify the calculation of amplitude, however, is not in our scope because it is merely a special case by cutting off our results when $m_\phi\lesssim 2m_e$. Nevertheless, we should point out that without using the third approximation we can use beam dump experiments to explore a larger parameter space.

As an example, we use the beam dump experiment E137 \cite{Bjorken:1988as} and the production of a new scalar boson, which we denote $\phi$. Interest in a new scalar boson arose recently because such particle which couples to standard model fermions can solve the proton radius puzzle and muonic anomalous magnetic moment discrepancy simultaneously \cite{TuckerSmith:2010ra,Liu:2016qwd}.

The outline of this paper is as follows. In Sec.~\ref{sec:dynamics}, we calculate the squared amplitude for 2 to 3 and 2 to 2 processes. In Sec. \ref{sec:cross section}, the cross sections for the 2 to 3 and 2 to 2 processes are calculated in the lab frame without any approximation. In Sec.~\ref{sec:WW approximation}, we introduce the WW approximation. In Sec.~\ref{sec:cross section comparison}, we derive and compare the cross sections with and without approximations. In Sec.~\ref{sec:particle production}, we compare the number of new particles produced in beam dump experiments with and without approximations. In Sec.~\ref{sec:data analysis}, we assume that this new scalar boson is observed and measured in beam dump experiment, determine the mass and coupling constant, and compare the results with and without approximations. A discussion is presented in Sec.~\ref{sec:discussion}.

\section{dynamics---a new scalar boson as an example}\label{sec:dynamics}
For simplicity, we assume that the new scalar boson $\phi$ only couples to electron by a Yukawa interaction, i.e. the scalar boson does not couple to other standard model fermions other than electron. The Lagrangian in the mostly-plus metric is
\begin{align}
\mathcal{L}\supset-\frac{1}{2}(\partial\phi)^2-\frac{1}{2}m_\phi^2\phi^2+e\epsilon\phi\bar\psi\psi
\end{align}
where $\epsilon=g/e$, $e$ is the electric charge, and $\psi$ is the electron field. Once the scalar boson is produced, it will decay to photons pairs through the electron loop,
\begin{align}\label{eq:decay to electrons}
\Gamma_{\phi\to\gamma\gamma}=\epsilon^2\frac{\alpha^3}{4\pi^2}\frac{m_\phi^3}{m_e^2}f\left(\frac{m_\phi^2}{4m_e^2}\right),
\end{align}
where $m_e$ is the electron mass and $f(\tau)=\frac{1}{4\tau^2}\left|1+(1-\frac{1}{\tau})\left(\sin^{-1}\sqrt{\tau}\right)^2\right|^2$. If $m_\phi>2m_e$, the scalar boson can also decay to electron pairs,
\begin{align}\label{eq:decay to photons}
\Gamma_{\phi\to e^+e^-}=\epsilon^2\frac{\alpha}{2}m_\phi\left(1-\frac{4m_e^2}{m_\phi^2}\right)^{3/2}.
\end{align}

\subsection{2 to 3 production}
\begin{figure}
\centering
\includegraphics[scale=0.8]{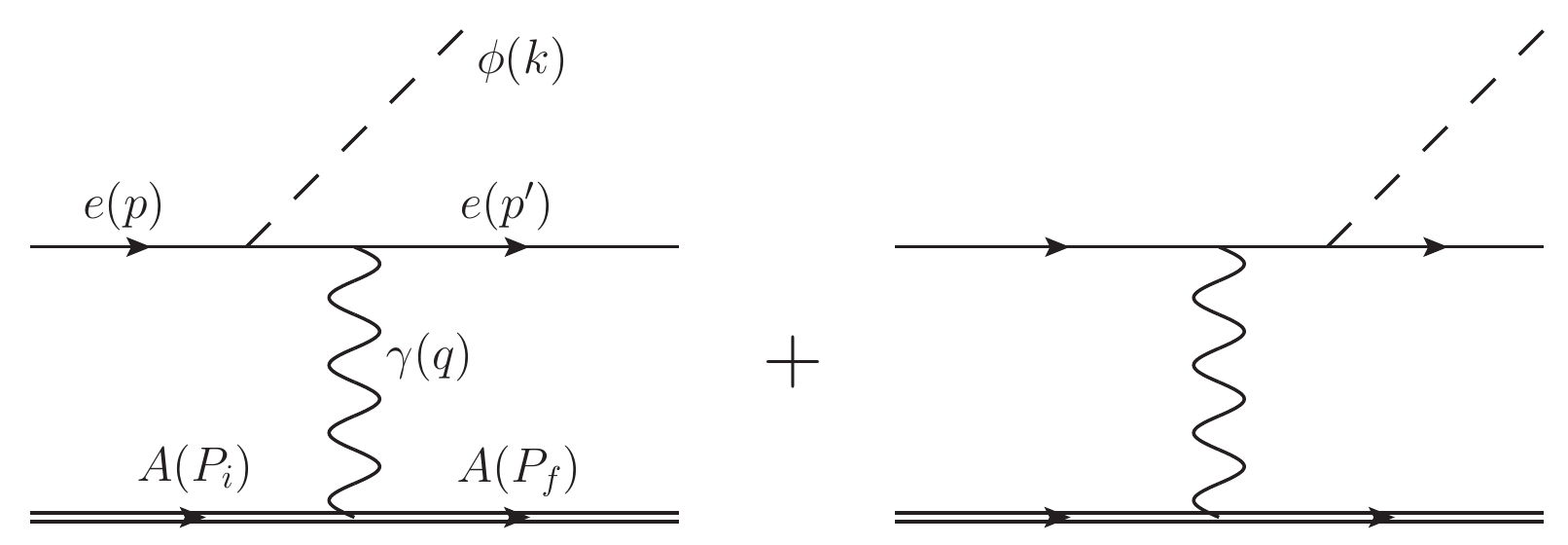}
\caption{\label{fig:2 to 3} Lowest order 2 to 3 production process: $e(p)+A(P_i)\rightarrow e(p')+A(P_f)+\phi(k)$. $A$, $\gamma$, $e$, and $\phi$ stand for the target atom, photon, electron, and the new scalar boson.}
\end{figure}

The leading production process is the bremsstrahlung-like radiation of the scalar from the electron, shown in Fig.~\ref{fig:2 to 3},
\begin{align}\label{eq:2 to 3 production process}
e(p)+A(P_i)\rightarrow e(p')+A(P_f)+\phi(k)
\end{align}
where $e$, $A$, and $\phi$ stand for electron, target atom, and the new scalar boson, respectively. We define the following quantities using the mostly-plus metric
\begin{align}\label{eq:2 to 3 variables}
\tilde{s}&=-(p'+k)^2-m_e^2=-2p'\cdotp k+m_\phi^2\nonumber\\
\tilde{u}&=-(p-k)^2-m_e^2=2p\cdotp k+m_\phi^2\nonumber\\
t_2&=-(p'-p)^2=2p'\cdotp p+2m_e^2\\
q&=P_i-P_f\nonumber\\
t&=q^2\nonumber
\end{align}
which satisfy
\begin{align}
\tilde{s}+t_2+\tilde{u}+t=m_\phi^2.
\end{align}

For definiteness, we assume the atom is a scalar boson (its spin is not consequential here) so that the Feynman rule for the photon-atom vertex is 
\begin{align}
ieF(q^2)(P_i+P_f)_\mu\equiv ieF(q^2)P_\mu
\end{align}
where $F(q^2)$ is the form factor which accounts for the nuclear form factor \cite{DeJager:1987qc} and the atomic form factor \cite{atomic form factor}. Here, we only include the elastic form factor since the contribution of the inelastic one is much smaller and can be neglected in computing the cross section. The amplitude of the process in Fig.~\ref{fig:2 to 3} is
\begin{align}
\mathcal{M}^{2\to3}&=e^2g\frac{F(q^2)}{q^2}\bar{u}_{p',s'}\left[\slashed{P}\frac{-(\slashed{p}-\slashed{k})+m_e}{-\tilde{u}}+\frac{-(\slashed{p'}+\slashed{k})+m_e}{-\tilde{s}}\slashed{P}\right]u_{p,s}
\end{align}
where $u_{p,s}$ is the electron spinor; $s$ and $s'$ are equal to $\pm 1$. After averaging and summing over initial and final spins, we have
\begin{align}
\overline{|\mathcal{M}^{2\to3}|^2}=\left(\frac{1}{2}\sum_s\right)\sum_{s'}|\mathcal{M}^{2\to3}|^2=e^4g^2\frac{F(q^2)^2}{q^4}\mathcal{A}^{2\to3}
\end{align}
where
\begin{align}
\mathcal{A}^{2\to3}=-\frac{(\tilde{s}+\tilde{u})^2}{\tilde{s}\tilde{u}}P^2-4\frac{t}{\tilde{s}\tilde{u}}(P\cdotp k)^2-\frac{(\tilde{s}+\tilde{u})^2}{\tilde{s}^2\tilde{u}^2}(m_\phi^2-4m_e^2)\left[P^2 t+4\left(\frac{\tilde{u}P\cdotp p+\tilde{s}P\cdotp p'}{\tilde{s}+\tilde{u}}\right)^2\right].
\end{align}

\subsection{2 to 2 production}
\begin{figure}
\centering
\includegraphics[scale=0.8]{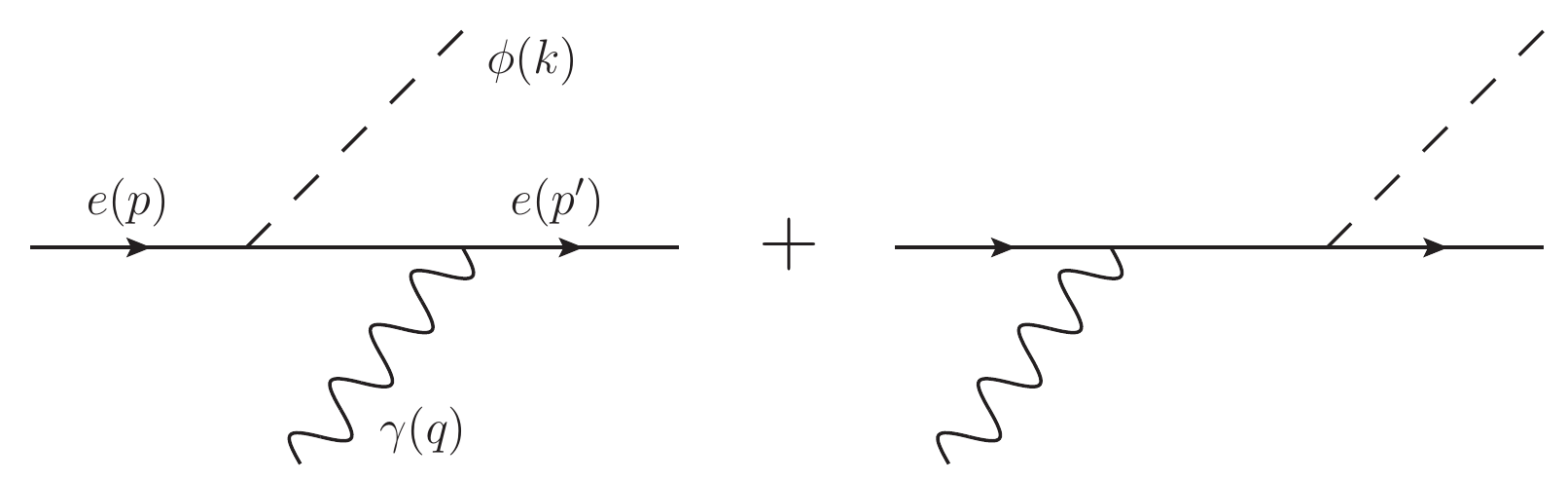}
\caption{\label{fig:2 to 2}  Lowest order 2 to 2 production process: $e(p)+\gamma(q)\rightarrow e(p')+\phi(k)$. $\gamma$, $e$, and $\phi$ stand for photon, electron, and the new scalar boson.}
\end{figure}

For the 2 to 2 process in Fig. \ref{fig:2 to 2}, a ``subprocess'' of the full 2 to 3 interaction,
\begin{align}\label{eq:2 to 2 production process}
e(p)+\gamma(q)\rightarrow e(p')+\phi(k).
\end{align} 
With the same definition in Eq. (\ref{eq:2 to 3 variables}), $\tilde{s}$, $\tilde{u}$, and $t_2$ satisfy
\begin{align}
\tilde{s}+t_2+\tilde{u}&=m_\phi^2
\end{align}
and the amplitude in Fig. \ref{fig:2 to 2} is
\begin{align}
\mathcal{M}^{2\to2}=eg\epsilon^\mu_\lambda\bar{u}_{p',s'}\left[\gamma_\mu\frac{-(\slashed{p}-\slashed{k})+m_e}{-\tilde{u}}+\frac{-(\slashed{p'}+\slashed{k})+m_e}{-\tilde{s}}\gamma_\mu\right]u_{p,s}
\end{align}
where $\epsilon$ is the photon polarization vector and $\lambda=\pm 1$. After averaging and summing over the initial and final spins and polarization,
\begin{align}\label{eq:2 to 2 M}
\overline{|\mathcal{M}^{2\to2}|^2}&=\left(\frac{1}{2}\sum_s\right)\sum_{s'}\left(\frac{1}{2}\sum_\lambda\right)|\mathcal{M}^{2\to2}|^2=e^2g^2\mathcal{A}^{2\to2}
\end{align}
where
\begin{align}\label{eq:2 to 2 A}
\mathcal{A}^{2\to2}=&-\frac{(\tilde{s}+\tilde{u})^2}{\tilde{s}\tilde{u}}+2(m_\phi^2-4m_e^2)\left[\left(\frac{\tilde{s}+\tilde{u}}{\tilde{s}\tilde{u}}\right)^2m_e^2-\frac{t_2}{\tilde{s}\tilde{u}}\right].
\end{align}

\section{cross section}\label{sec:cross section}
\subsection{2 to 3}
The cross section for the 2 to 3 process, see Eq. (\ref{eq:2 to 3 production process}) and Fig. \ref{fig:2 to 3}, in the lab frame is given by
\begin{align}
d\sigma=\frac{1}{4|\textbf{p}|M}\overline{|\mathcal{M}^{2\to3}|^2}(2\pi)^4\delta^4(p'+k-p-q)\frac{d^3\textbf{p}'}{(2\pi)^3 2E'}\frac{d^3\textbf{P}_f}{(2\pi)^3 2E_f}\frac{d^3\textbf{k}}{(2\pi)^3 2E_k}
\end{align}
where $M$ is the mass of the target atom. Integrating over $\mathbf{p}'$ and changing the variable from $\mathbf{P}_f$ to $\mathbf{q}$, we have
\begin{align}
d\sigma=\frac{\overline{|\mathcal{M}^{2\to3}|^2}}{1024\pi^5|\textbf{p}|M E_f E' E_k}\delta(E'+E_k-E-q_0)d^3\textbf{q}d^3\textbf{k}.
\end{align}
In order to integrate over $\mathbf{q}$, we choose the spherical coordinate $(Q,\theta_q,\phi_q)$ where $Q=|\mathbf{q}|$, and $\theta_q$ and $\phi_q$ are the polar and azimuthal angles of \textbf{q} in the direction of $\mathbf{V}=\mathbf{k}-\mathbf{p}$. First, we use the remaining $\delta$-function to integrate out $Q$, and then change variables from $\theta_q$ to $t$. We obtain
\begin{align}
d\sigma=\frac{d^3\textbf{k}}{128\pi^4|\textbf{p}|V E_k}\int^{t_{max}}_{t_{min}}dt\left(\frac{1}{8M^2}\int_0^{2\pi}\frac{d\phi_q}{2\pi}\overline{|\mathcal{M}^{2\to3}|^2}\right)
\end{align}
where $V=|\textbf{V}|$, $t(Q)=q^2=2M(\sqrt{M^2+Q^2}-M)$, $t_{max}=t(Q_+)$, $t_{min}=t(Q_-)$, and
\begin{align}
Q_\pm=\frac{V[\tilde{u}+2M(E'+E_f)]\pm(E'+E_f)\sqrt{\tilde{u}^2+4M\tilde{u}(E'+E_f)+4M^2V^2}}{2(E'+E_f)^2-2V^2}.
\end{align}
Integrate over the polar angle, $\theta$, and azimuthal angle of \textbf{k} in the direction of \textbf{p}, and then change the variable from $|\mathbf{k}|$ to $x$ where $x\equiv E_k/E$. We have
\begin{align}\label{eq:2 to 3}
\frac{d\sigma}{dx d\cos\theta}&=\frac{|\textbf{k}|E}{64\pi^3|\textbf{p}|V}\int^{t_{max}}_{t_{min}}dt\left(\frac{1}{8M^2}\int_0^{2\pi}\frac{d\phi_q}{2\pi}\overline{|\mathcal{M}^{2\to3}|^2}\right)\nonumber\\
&=\epsilon^2\alpha^3\frac{|\textbf{k}|E}{|\textbf{p}|V}\int^{t_{max}}_{t_{min}}dt\frac{F(t)^2}{t^2}\left(\frac{1}{8M^2}\int_0^{2\pi}\frac{d\phi_q}{2\pi}\mathcal{A}^{2\to3}\right).
\end{align}

\subsection{2 to 2}
The 2 to 2 cross section, see Eq. (\ref{eq:2 to 2 production process}) and Fig. \ref{fig:2 to 2}, in the lab frame is straightforwardly expressed in terms of the amplitude,
\begin{align}\label{eq:2 to 2}
\frac{d\sigma}{d(p\cdotp k)}=2\frac{d\sigma}{dt_2}=\frac{\overline{|\mathcal{M}^{2\to2}|^2}}{8\pi\tilde{s}^2}=\epsilon^2\alpha^2\frac{2\pi}{\tilde{s}^2}\mathcal{A}^{2\to2}.
\end{align}

\section{Weizs\"{a}cker-Williams approximation}\label{sec:WW approximation}
It is explained in Ref.~\cite{Kim:1973he} that the WW approximation relies on the incoming electron energy being much greater than $m_\phi$ and $m_e$, such that the final state electron and scalar boson are highly collinear. In that case the phase space integral can be approximated by
\begin{align}\label{eq:WW}
\frac{1}{8M^2}\int\frac{d\phi_q}{2\pi}\mathcal{A}^{2\to3}\approx\frac{t-t_{min}}{2t_{min}}\mathcal{A}^{2\to2}_{t=t_{min}}.
\end{align}
With the WW approximation, Eq.~(\ref{eq:2 to 3}) can be approximated to be
\begin{align}
\frac{d\sigma}{dx d\cos\theta}\approx\epsilon^2\alpha^3\frac{|\textbf{k}|E}{|\textbf{p}|V}\frac{\mathcal{A}^{2\to2}_{t=t_{min}}}{2t_{min}}\chi
,\end{align}
where
\begin{align}\label{eq:chi}
\chi=\int^{t_{max}}_{t_{min}}dt\frac{t-t_{min}}{t^2}F(t)^2.
\end{align}
Using Eq. (\ref{eq:2 to 2}), we have
\begin{align}
\frac{d\sigma}{dx d\cos\theta}\approx\frac{\alpha\chi}{4\pi}\frac{|\textbf{k}|E}{|\textbf{p}|V}\frac{\tilde{s}^2}{t_{min}}\left.\frac{d\sigma}{d(p\cdotp k)}\right|_{t=t_{min}}.
\end{align}
Following the discussion in Refs.~\cite{Bjorken:2009mm,Tsai:1986tx}, near $t=t_{min}$ (when $\mathbf{q}$ and $\mathbf{V}=\mathbf{k}-\mathbf{p}$ are collinear), we can approximate the following quantities
\begin{align}\label{eq:WW variables}
\tilde{s}&\approx-\frac{\tilde{u}}{1-x}\nonumber\\
\tilde{u}&\approx-xE^2\theta_\phi^2-m_\phi^2\frac{1-x}{x}-m_e^2 x\nonumber\\
t_2&\approx\frac{\tilde{u}x}{1-x}+m_\phi^2\\
V&\approx E(1-x)\nonumber\\
t_{min}&\approx\frac{\tilde{s}^2}{4E^2}\nonumber
\end{align}
Using Eq. (\ref{eq:WW variables}), we arrive at the well-known equation \cite{Bjorken:2009mm,Tsai:1986tx}
\begin{align}\label{eq:WW Tsai}
\frac{d\sigma}{dx d\cos\theta}\approx\frac{\alpha\chi}{\pi}\frac{xE^2\beta}{1-x}\left.\frac{d\sigma}{d(p\cdotp k)}\right|_{t=t_{min}}
\end{align}
where $\beta=\sqrt{1-m_\phi^2/E_k^2}$. Note that in Eq. (\ref{eq:WW Tsai}) $d\sigma/d(p\cdotp k)$ is evaluated at $t=t_{min}$. So the amplitude $\mathcal{A}^{2\to2}$ in Eq. (\ref{eq:2 to 2}) evaluated at $t=t_{min}$ using Eq. (\ref{eq:WW variables}) is
\begin{align}\label{eq:2 to 2 A tmin}
\mathcal{A}^{2\to2}_{t=t_{min}}\approx\frac{x^2}{1-x}+2(m_\phi^2-4m_e^2)\frac{\tilde{u}x+m_\phi^2(1-x)+m_e^2x^2}{\tilde{u}^2}.
\end{align}

\section{cross section comparison}\label{sec:cross section comparison}

\begin{figure}
\centering
\subfigure[\;$d\sigma/(\epsilon^2dx)$]{\includegraphics[scale=0.95]{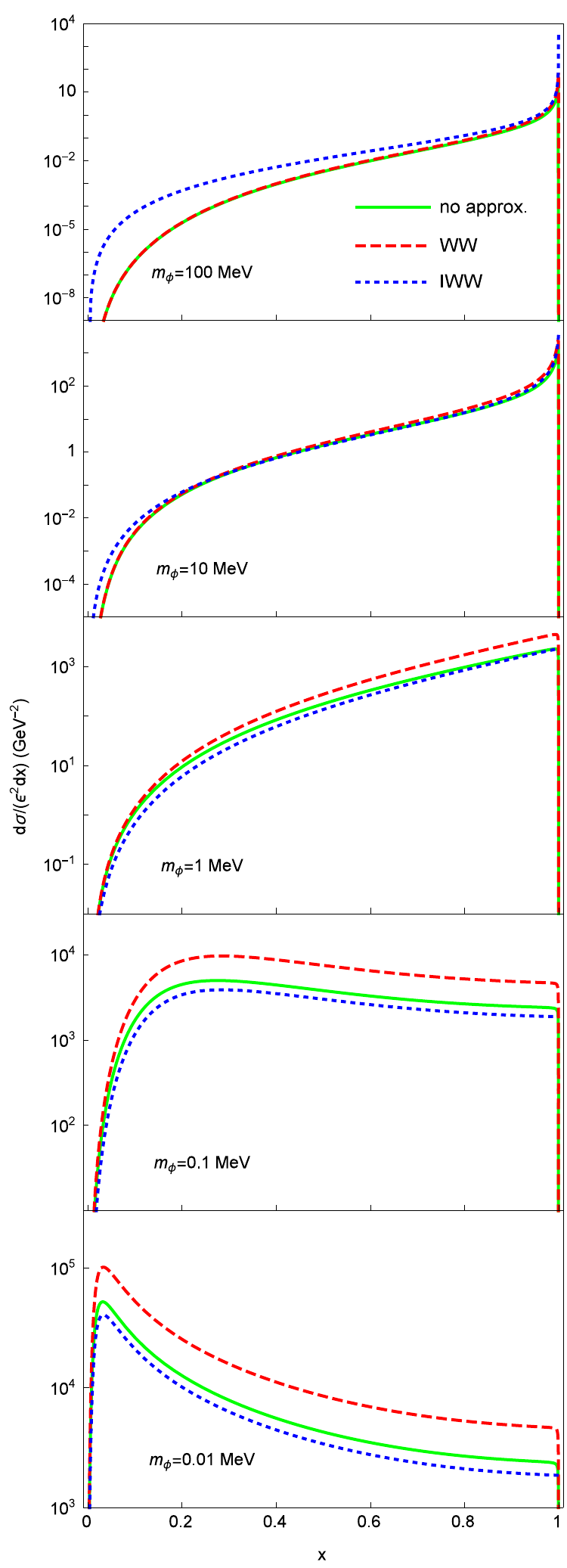}}
\subfigure[\;relative error of $d\sigma/(\epsilon^2 dx)$]{\includegraphics[scale=0.95]{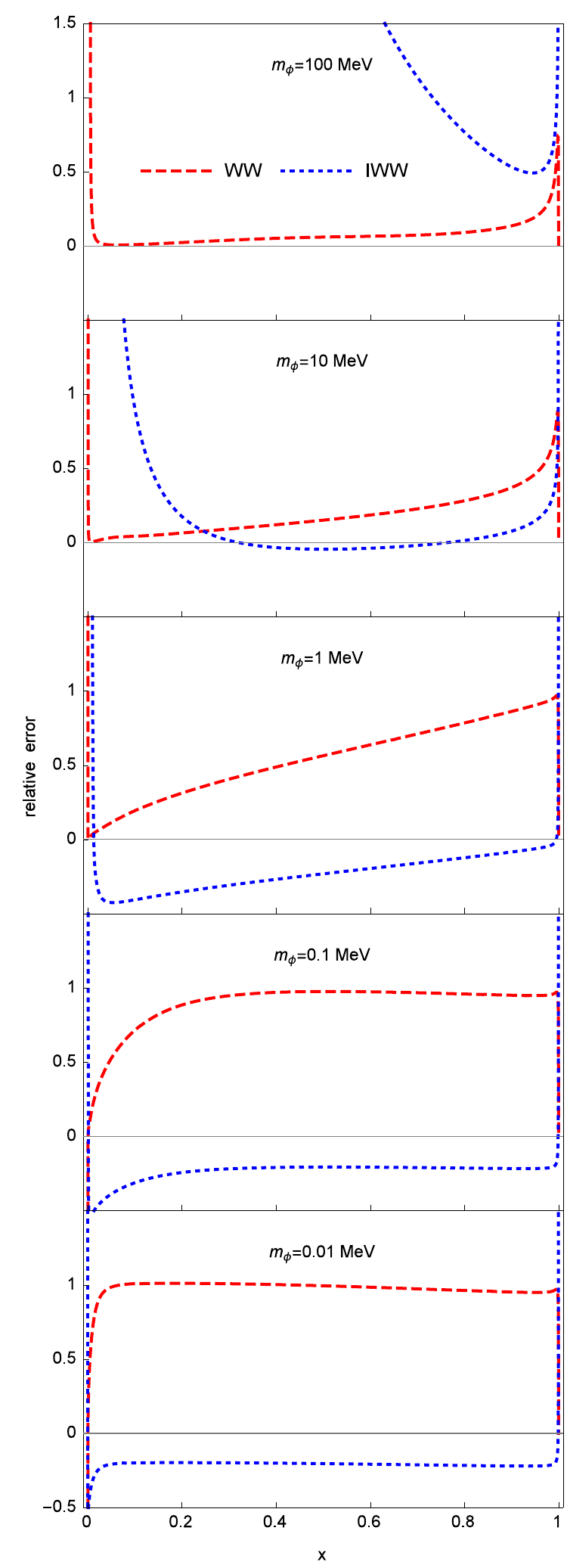}}
\caption{\label{fig:cross_section} The solid green, dashed red, and dotted blue lines correspond to the differential cross section with no, WW, and IWW approximation. The relative error of $\mathcal{O}$ is defined by $(\mathcal{O}_{\rm approx.}-\mathcal{O}_{\rm exact})/\mathcal{O}_{\rm exact}$.}
\end{figure}

To test approximations of the cross section for $\phi$ production, we examine three cases. 
\begin{enumerate}
\item The complete calculation, Eq. (\ref{eq:2 to 3}),
\begin{align}\label{eq:d sigma dx 1}
\frac{d\sigma}{dx}=\epsilon^2\alpha^3\frac{|\textbf{k}|E}{|\textbf{p}|}\int_0^{\theta_{max}} d\cos\theta\frac{1}{V}\int^{t_{max}}_{t_{min}}dt\frac{F(t)^2}{t^2}\left(\frac{1}{8M^2}\int_0^{2\pi}\frac{d\phi_q}{2\pi}\mathcal{A}^{2\to3}\right)
\end{align}
where $\theta_{max}$ depends on the configuration of the detector. For beam dump E137, $\theta_{max}\approx 4.4\times10^{-3}$.

\item WW: using the WW approximation, Eq. (\ref{eq:WW}),
\begin{align}\label{eq:d sigma dx 2}
\left(\frac{d\sigma}{dx}\right)_{WW}=2\epsilon^2\alpha^3|\textbf{k}|E(1-x)\int_0^{\theta_{max}} d\cos\theta\frac{\mathcal{A}^{2\to2}_{t=t_{min}}}{\tilde{u}^2}\chi
\end{align}
where $\theta_{max}$ is the same as the first case and $\chi$ is defined in Eq. (\ref{eq:chi}). Note that the upper and lower limits of $\chi$ depend on $x$ and $\theta$.

\item Improved WW (IWW): If the upper and lower limits of the $t$-integral in $\chi$ in Eq. (\ref{eq:d sigma dx 2}) are not sensitive to $x$ and $\theta$; i.e., the integration limit can be set to be independent of $x$ and $\theta$, we can further approximate the integration limits of $t$. Similar to the argument in Ref.~\cite{Bjorken:2009mm}, we set
\begin{align}\label{eq:tmin tmax}
t_{min}=\left(\frac{m_\phi^2}{2E}\right)^2 {\rm\; and\;\;} t_{max}=m_\phi^2+m_e^2
\end{align}
which is valid when the production cross section is dominantly collinear with $x$ close to 1. The difference in $t_{max}$ between \cite{Bjorken:2009mm} and our approach is because we do not assume $m_\phi\gg m_e$. Therefore, we can pull $\chi$ out of the integral over $\cos\theta$. Then, changing variables from $\cos\theta$ to $\tilde{u}$ and extending the lower limit of $\tilde{u}$ to $-\infty$, we have
\begin{align}\label{eq:d sigma dx 3}
\left(\frac{d\sigma}{dx}\right)_{IWW}&=\epsilon^2\alpha^3\chi\frac{|\textbf{k}|}{E}\frac{1-x}{x}\int^{\tilde{u}_{max}}_{-\infty}d\tilde{u}\frac{\mathcal{A}^{2\to2}_{t=t_{min}}}{\tilde{u}^2}\\
&=\epsilon^2\alpha^3\chi\frac{|\textbf{k}|}{E}\frac{m_e^2(2-x)^2-2x \tilde{u}_{max}}{3\tilde{u}_{max}^2}
\end{align}
where $\tilde{u}_{max}=-m_\phi^2\frac{1-x}{x}-m_e^2 x$ and in the last line, we use Eq. (\ref{eq:2 to 2 A tmin}). We emphasize that the name ``improved" means reducing the computational time (because of one fewer integral than in the WW approximation above) and does not imply more accuracy.

\end{enumerate}

In Fig. \ref{fig:cross_section}, we show the cross sections in each of the three cases for five values of the scalar boson mass, setting the incoming electron beam energy to 20 GeV. In both approximations, the cross section is of the same order of magnitude as that using the complete calculation. However, there are regions where there are ${\cal O}\left(1\right)$ relative errors. The WW approximation (dashed red lines in Fig.~\ref{fig:cross_section}) can differ from the complete calculation by 100\% when $m_\phi\lesssim 1$ MeV; in the IWW case (dotted blue lines in Fig.~\ref{fig:cross_section}), the approximation starts to fail when $m_\phi\gtrsim 100$ MeV.

\section{particle production}\label{sec:particle production}

There are two characteristic lengths which are crucial in beam dump experiments. The first is the decay length of the new particle in the lab frame,
\begin{align}\label{eq:absorption process}
l_\phi=\frac{E_k}{m_\phi}\frac{1}{\Gamma_\phi},
\end{align}
where $\Gamma_\phi=\Gamma_{\phi\to e^+e^-}+\Gamma_{\phi\to\gamma\gamma}$, see Eq. (\ref{eq:decay to electrons}) and (\ref{eq:decay to photons}). The new particle, after production, must decay after going through the target and shielding and before going through the detector in order to be observed. If the target is thick (much greater than a radiation length), most of the new particles will be produced in the first few radiation lengths. The production rate is approximately proportional to the probability $e^{-L_{sh}/l_\phi}(1-e^{-L_{dec}/l_\phi})$, where $L_{sh}$ is length of the target and shield and $L_{dec}$ is length for the new particle to decay into electron or photon pairs after the shield and before the detector.

The second characteristic length is the absorption length
\begin{align}
\lambda=\frac{1}{n_e\sigma_{abs}},
\end{align}
where $n_e$ is the number density of the target electrons and $\sigma_{abs}$ is the cross section of absorption process. The leading process of absorption is
\begin{align}\label{eq:absorption}
e(p)+\phi(k)\rightarrow e(p')+\gamma(q),
\end{align} 
which is related to the 2 to 2 production process Eq. (\ref{eq:2 to 2 production process}) via crossing symmetry $\tilde{s}\leftrightarrow\tilde{u}$. Since Eq. (\ref{eq:2 to 2 A}) is symmetric in $\tilde{s}\leftrightarrow\tilde{u}$, the algebraic form of amplitude squared of absorption process is the same as Eq. (\ref{eq:2 to 2 A}) but differs by a factor 2 from summing over final state instead of averaging over initial state in Eq. (\ref{eq:2 to 2 M})
\begin{align}
\mathcal{A}^{2\to2}_{abs}=2\mathcal{A}^{2\to2}.
\end{align}
The cross section of the process (\ref{eq:absorption process}) is 
\begin{align}
\frac{d\sigma}{d\Omega}&=\frac{1}{64\pi^2 m_e}\frac{|\mathbf{q}|}{|\mathbf{k}|}\frac{\overline{|\mathcal{M}^{2\to2}_{abs}|^2}}{E_k+m_e-|\mathbf{k}|\cos\theta_\gamma}\\
\sigma_{abs}&=\frac{\pi\epsilon^2\alpha^2}{m_e|\mathbf{k}|}\int_{-1}^1 d\cos\theta_\gamma\frac{|\mathbf{q}|\mathcal{A}^{2\to2}}{E_k+m_e-|\mathbf{k}|\cos\theta_\gamma},
\end{align}
where $\theta_\gamma$ is the angle between outgoing photon and incoming new particle. The new particle, after produced, must not be absorbed by the target and shield to be detected. If the target is thick (much greater than absorption length), the production rate will be approximately proportional to the probability $e^{-L_{sh}/\lambda}$.

The number of the new particles produced in terms of the cross section (without considering the absorption process) can be found in, {\it e.g.}, Refs.~\cite{Bjorken:2009mm,Tsai:1986tx,Andreas:2012mt}. Using the thick target approximation and including the absorption process, we find
\begin{align}
N_\phi\approx\frac{N_eX}{M}\int_{E_{min}}^{E_0}dE\int_{x_{min}}^{x_{max}}dx\int_0^TdtI_e(E_0,E,t)\frac{d\sigma}{dx}e^{-L_{sh}\left(\frac{1}{l_\phi}+\frac{1}{\lambda}\right)}(1-e^{-L_{dec}/l_\phi}),
\end{align}
where $M$ is the mass of the target atom (aluminium); $N_e$ is the number of incident electrons; $X$ is the unit radiation length of the target; $E_0$ is the incoming electron beam energy, $E_{min}=m_e+\max(m_\phi,E_{cut})$ and $x_{min}=\frac{\max(m_\phi,E_{cut})}{E}$ where $E_{cut}$ is the measured energy cutoff depending on the detectors; $x_{max}$, which is smaller  but very close to 1 ($x_{max}$ can be approximated to be $1-\frac{m_e}{E}$ if the new particle and electron initial and final state are collinear); $T=\rho L_{sh}/X$ where $\rho$ is the density of the target; $l_\phi$ is the decay length of the new particle in lab frame; $\lambda$ is the absorption length of the new particle passing through the target and shield; $I_e$, derived in Ref.~\cite{Tsai:1966js}, is the energy distribution of the electrons after passing through a medium of $t$ radiation length  
\begin{align}
I_e(E_0,E,t)=\frac{\left(\ln\frac{E_0}{E}\right)^{bt-1}}{E_0\Gamma(bt)},
\end{align}
where $\Gamma$ is the gamma function and $b=4/3$. For beam dump E137 which we take as our prototypical setup, $E_0=20$ GeV and $E_{cut}=2$ GeV; $N_e=1.87\times 10^{{20}}$; $L_{sh}=179$ m and $L_{dec}=204$ m. The experiment has a null result which translates to 95\% C.L. of $N_\phi$ to be 3 events.

\begin{figure}
\centering
\subfigure[\;exclusion plot]{\includegraphics[scale=1.02]{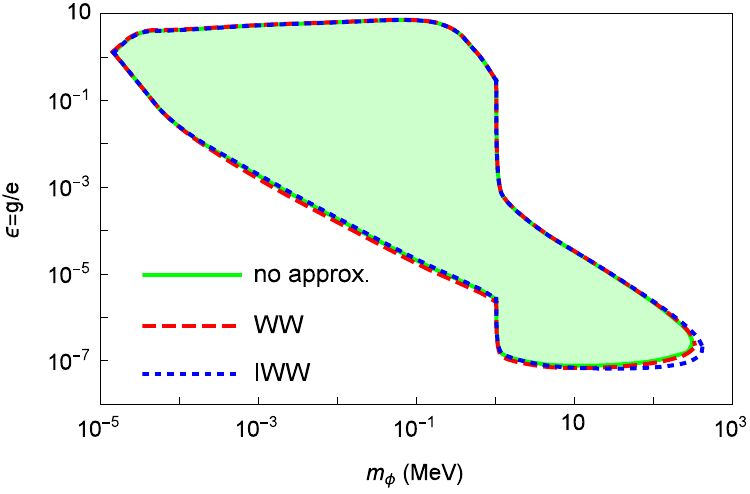}}
\subfigure[\;exclusion plot (zoomed in)]{\includegraphics[scale=1]{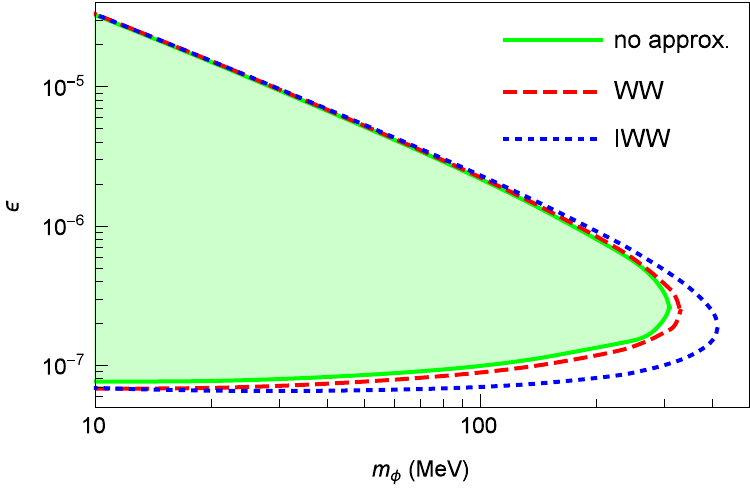}}
\subfigure[\;exclusion plot (linear scale)]{\includegraphics[scale=1]{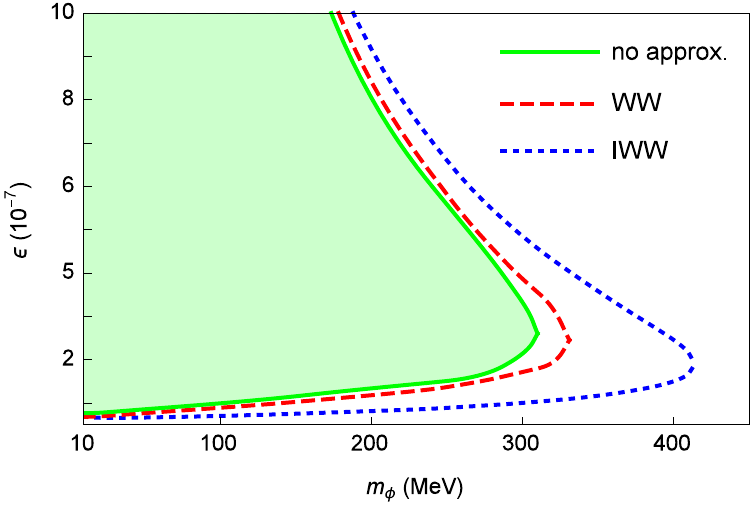}}
\subfigure[\;relative error of exclusion boundary]{\includegraphics[scale=1.03]{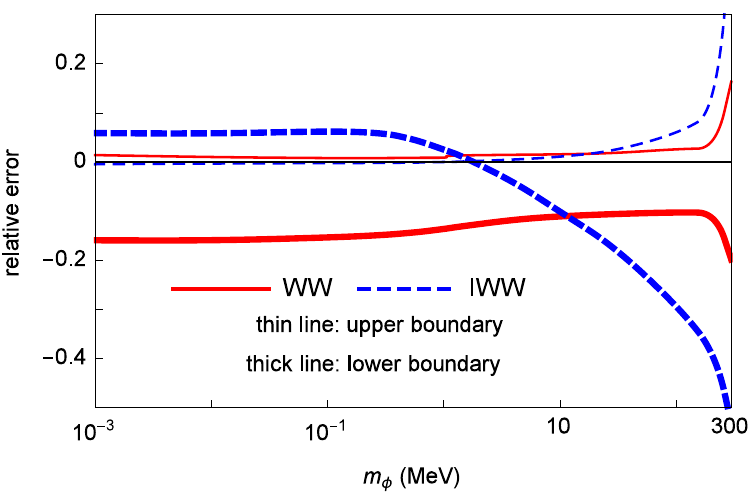}}
\caption{\label{fig:E137} (a)--(c) Exclusion (shaded region) plot for $\epsilon$ using the beam dump experiment E137. The solid green, dashed red, and dotted blue lines correspond to the differential cross section with no, WW, and IWW approximation. (d) The solid red and dashed blue lines correspond to the relative error of exclusion boundary with WW and IWW approximation. The thin and thick lines correspond to the upper and lower boundaries of the exclusion plot.}
\end{figure}

In Fig.~\ref{fig:E137}, we show regions of coupling and mass excluded by the lack of a signal at E137, using the three different ways to calculate the differential cross section, $d\sigma/dx$. Because of the exponential factor from decay and absorption lengths, the error in the exclusion plot due to making approximations to the cross section is smaller along the upper boundary, which is mainly determined by whether $\phi$ lives long enough to make it to the detector. With the WW approximation, the 100\% error in cross section causes an error of less than 20\% along the lower boundary, and in a log-log plot across several scales, a 20\% error is almost indistinguishable by eyesight. On the other hand, with the IWW approximation, the difference is clearly visible when $m_\phi\gtrsim 100$ MeV. We emphasize that the similarity of the exclusions with or without the approximations in a log-log plot means that the cross section approximations are good to the order of magnitude but the relative error can deviate at the ${\cal O}\left(1\right)$ level.

In Fig. \ref{fig:E137}, we see that the absorption process, Eq. (\ref{eq:absorption}), cuts off the exclusion plot around $\epsilon\sim\mathcal{O}(1)$ where the coupling of $\phi$ to electrons is of same order of the electromagnetic coupling. Therefore, in this region, there is another significant process to consider for beam dump experiments. This is the trapping process due to the rescattering
\begin{align}
e(p)+\phi(k)\rightarrow e(p')+\phi(k').
\end{align} 
The trapping process is expected to be as important as the absorption process in this example (new scalar particle, beam dump E137), and also cuts off the exclusion plot around $\epsilon\sim\mathcal{O}(1)$. However, in Fig. \ref{fig:E137} the region where $\epsilon>10^{-3}$  has been excluded by other experiments, such as electron $g-2$ \cite{Pospelov:2008zw,Bouchendira:2010es} and hydrogen Lamb shift \cite{Eides:2000xc}, which are discussed in Ref.~\cite{Liu:2016qwd} as well as astrophysical processes \cite{Essig:2013lka}. Therefore we do not include the trapping process in this example, but it might be crucial for other experiments.

\section{A positive signal}\label{sec:data analysis}
To further explore the accuracy of the approximations to the cross section, let us imagine that there is a signal of a new particle being produced at a beam dump experiment. In such a case, the mass and the coupling of this particle can be determined by examining the data, {\it i.e.}, the distribution of events as a function of energy deposited in the detector. We perform 3 sets of pseudoexperiment by using the setup of E137; assume that the scalar boson exists with $(m_\phi,\epsilon)=(110{\rm\;MeV,10^{-7}})$, $(m_\phi,\epsilon)=(200{\rm\;MeV,1.3\times10^{-7}})$, and $(m_\phi,\epsilon)=(0.3{\rm\;MeV,8\times 10^{-6}})$, which are outside of the current exclusion in Fig. \ref{fig:E137}. We increase the incoming beam luminosities by 36, 36, and 137 times (increasing the total number of electrons dumped into the target), so that the expected total number of events is around 100, 100, and 400. We assume that the resolution of the detector is 1 GeV (which means that there are 18 bins) and generate the ``observed" number of events using a Poisson distribution with the mean value from the complete calculation for each bin. Finally, we can fit the ``observed" data with the calculation with no, WW, and IWW approximation using $\chi^2$ test, and we assume that the variance of the calculated value also satisfies Poisson distribution ({\it i.e.} we ignore systematic errors on the observed numbers of events for simplicity). Therefore, the definition of $\chi^2$ becomes
\begin{align}
\chi^2=\sum_i\frac{(N_{cal,i}-N_{obs,i})^2}{\sigma^2_i}=\frac{(N_{cal,i}-N_{obs,i})^2}{N_{cal,i}+N_{obs,i}}
\end{align}
where $N_{cal}$ and $N_{obs}$ are calculated and ``observed" number of events; the subscript $i$ is for the bins. Since there are two independent parameters (mass and coupling) to fit, the $1\sigma$ and $2\sigma$ range correspond to $\Delta\chi^2=2.30$ and $\Delta\chi^2=6.18$, where $\Delta\chi^2=\chi^2-\chi^2_{min}$.

\begin{figure}
\centering
\subfigure[\;generated data]{\includegraphics[scale=1]{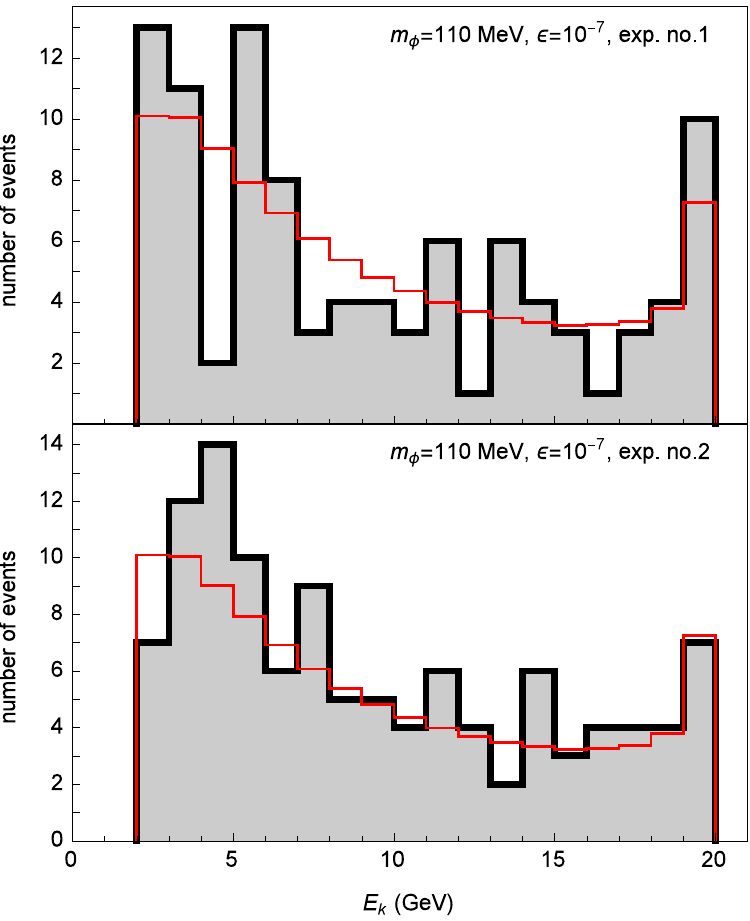}}
\subfigure[\;$\chi^2$ fit]{\includegraphics[scale=1.03]{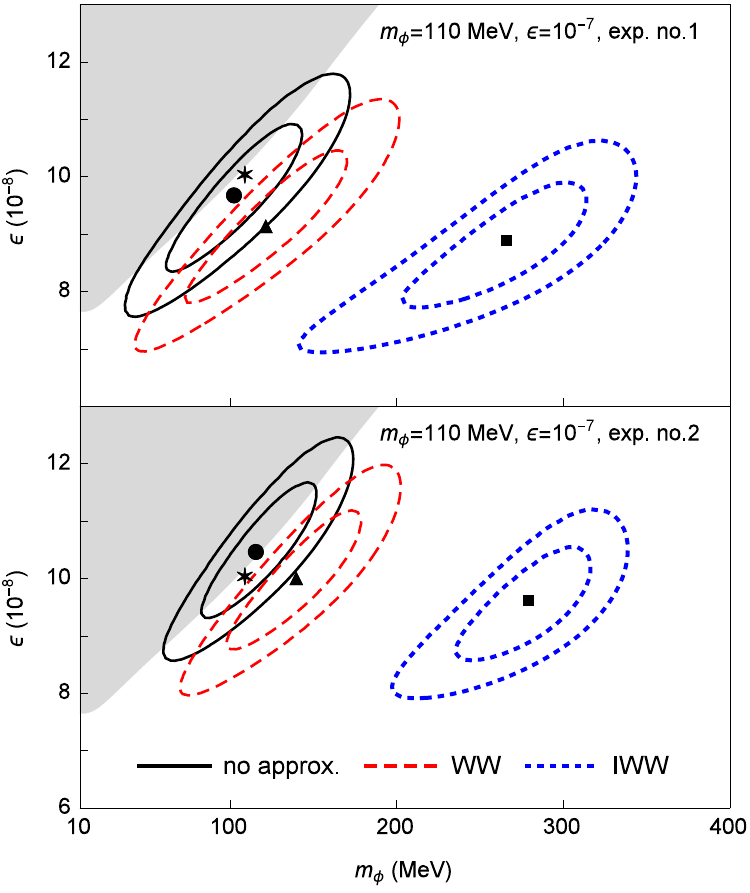}}
\caption{\label{fig:pseudo_exp_1} Assuming the scalar boson exists with $(m_\phi,\epsilon)=(110{\rm\;MeV,10^{-7}})$ and is observed in E137 with 36 times luminosity. (a) The number of events distribution with respect to the energy of the scalar boson: the thin red line is obtained by the complete calculation (no approximation), and the thick black lines is the ``data" generated by Poisson distribution with mean value given by the complete calculation. (b) The best fit point, $1\sigma$ range, and $2\sigma$ range with no, WW, and IWW approximation: the star is the ``true" value; the circle, triangle, and squares are the best fit parameters with no, WW, and IWW approximation, respectively; the black, dashed red, and dotted blue inner (outer) loop correspond to the $1\sigma$ ($2\sigma$) range with no, WW, and IWW approximation, respectively; the shaded area is the excluded region with no approximation from Fig. \ref{fig:E137}. The top and bottom rows correspond to the results of two separate pseudoexperiments.}
\end{figure}

\begin{figure}
\centering
\subfigure[\;generated data]{\includegraphics[scale=1]{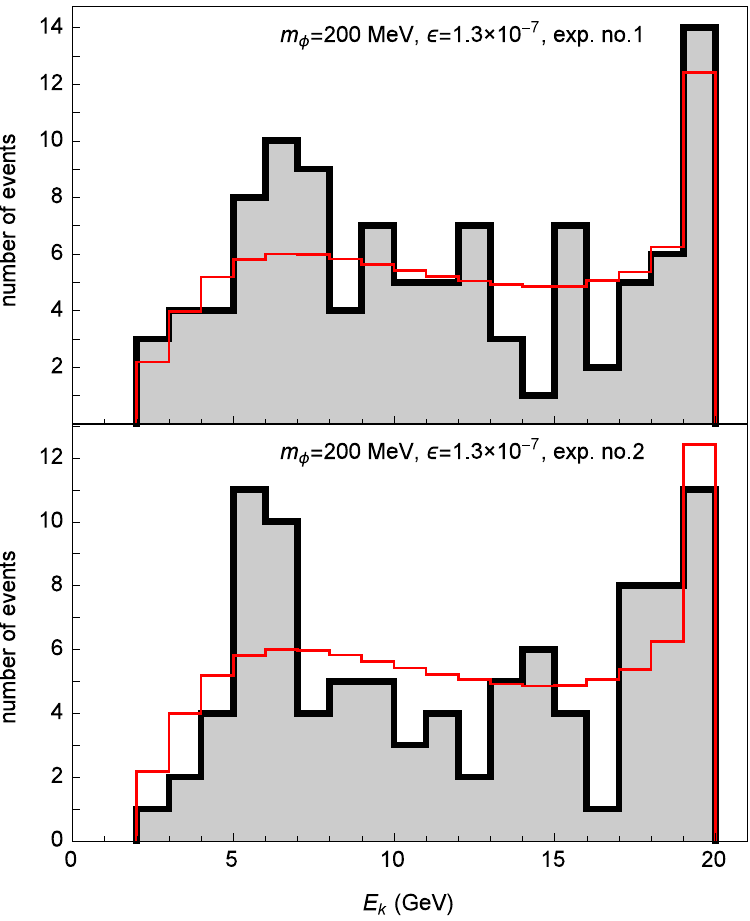}}
\subfigure[\;$\chi^2$ fit]{\includegraphics[scale=1]{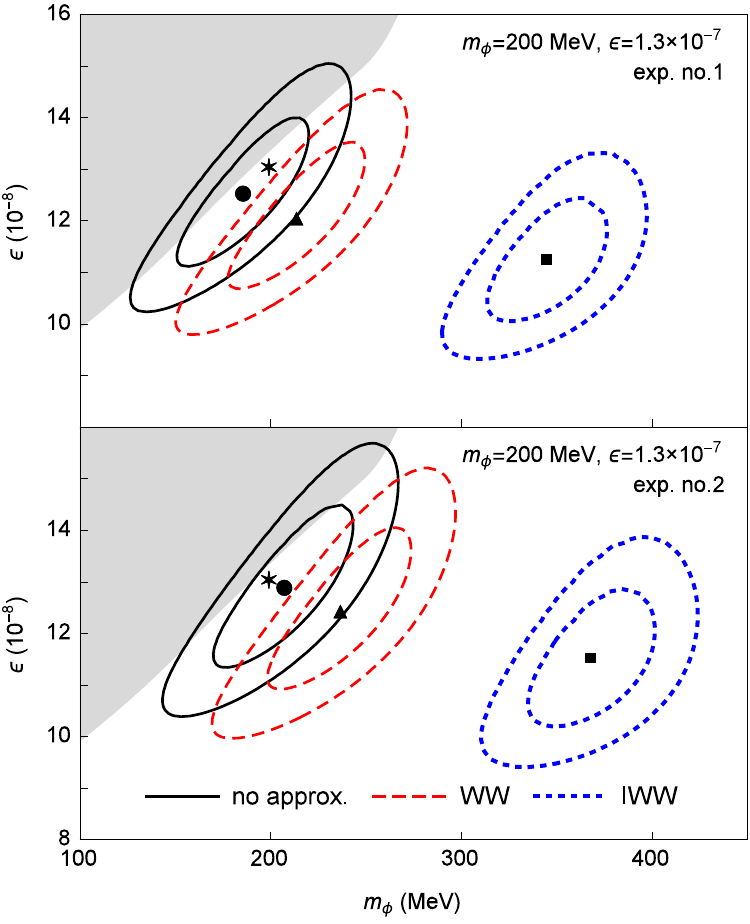}}
\caption{\label{fig:pseudo_exp_2} Assuming the scalar boson exists with $(m_\phi,\epsilon)=(200{\rm\;MeV,1.3\times10^{-7}})$ and is observed in E137 with 36 times luminosity. See the caption in Fig. \ref{fig:pseudo_exp_1} for details.}
\end{figure}

\begin{figure}
\centering
\subfigure[\;generated data]{\includegraphics[scale=0.75]{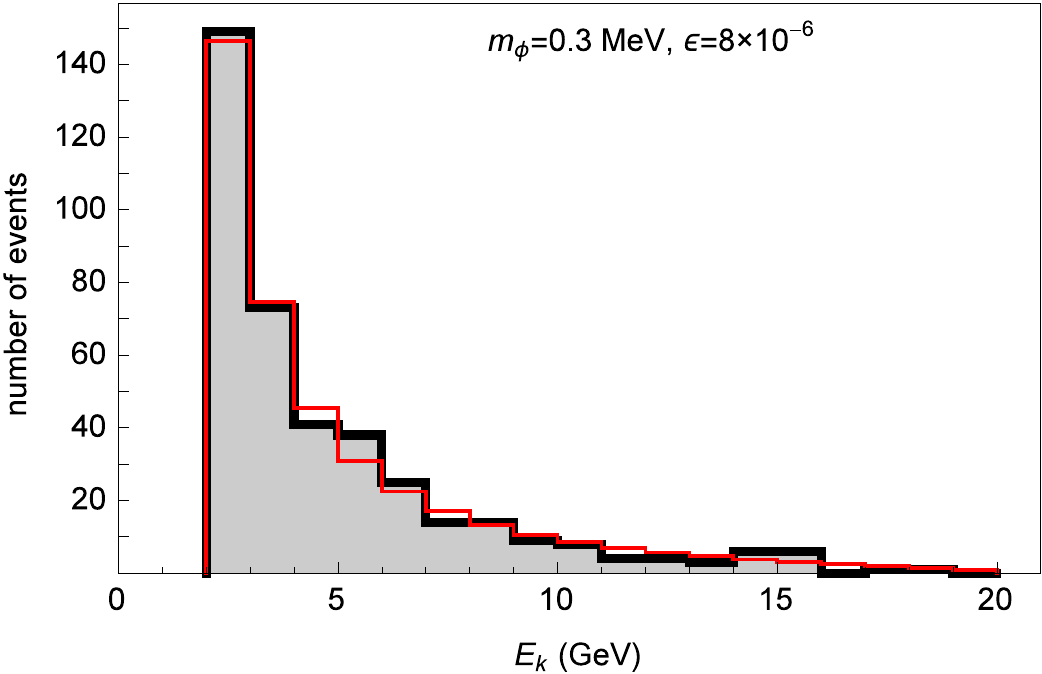}}
\subfigure[\;$\chi^2$ fit]{\includegraphics[scale=0.75]{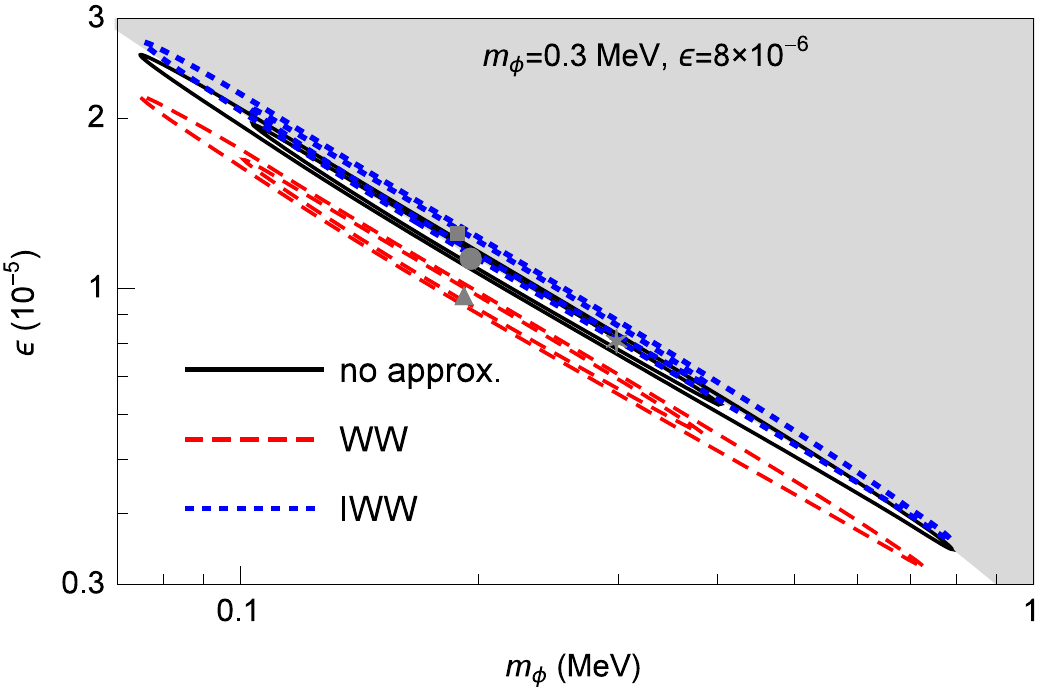}}
\subfigure[\;$\chi^2$ fit (zoomed in)]{\includegraphics[scale=0.75]{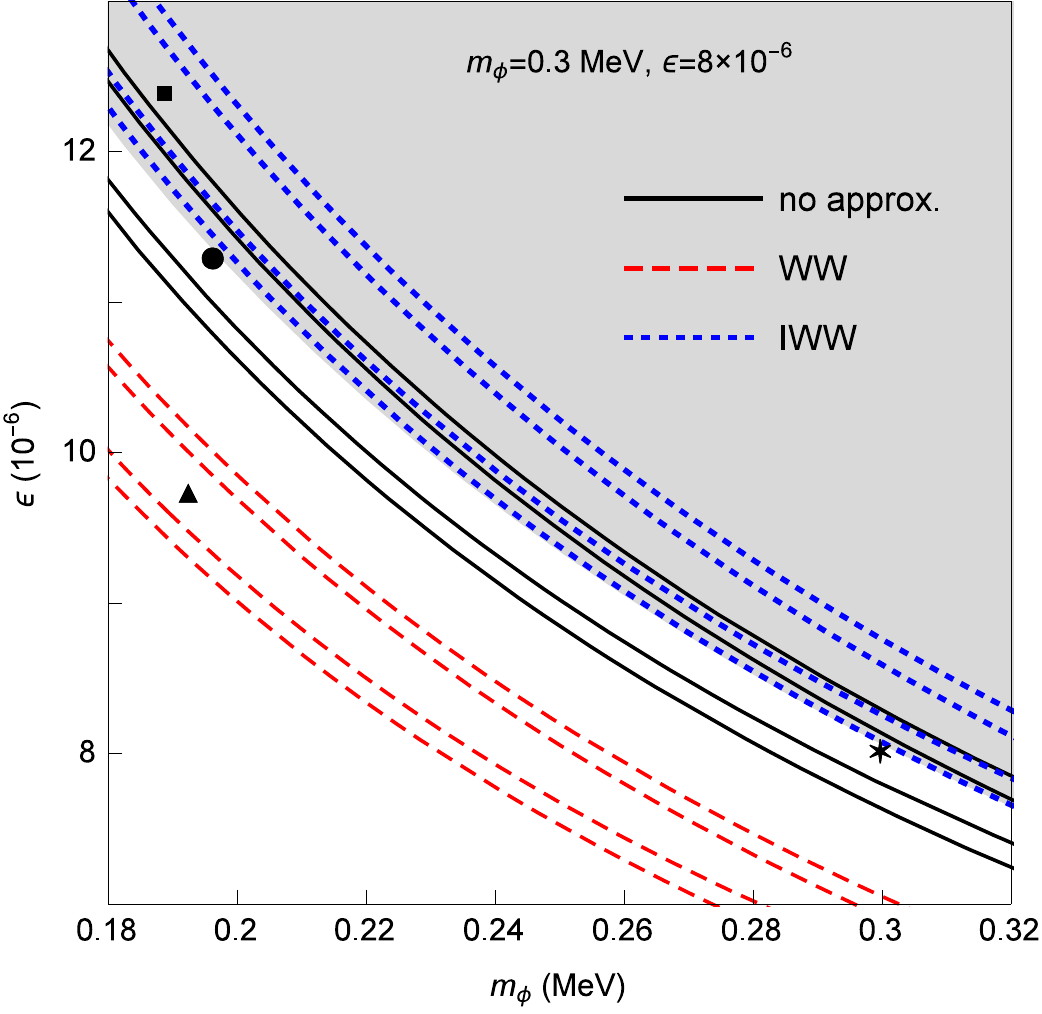}}
\subfigure[\;$\chi^2$ fit (change of coordinate)]{\includegraphics[scale=0.76]{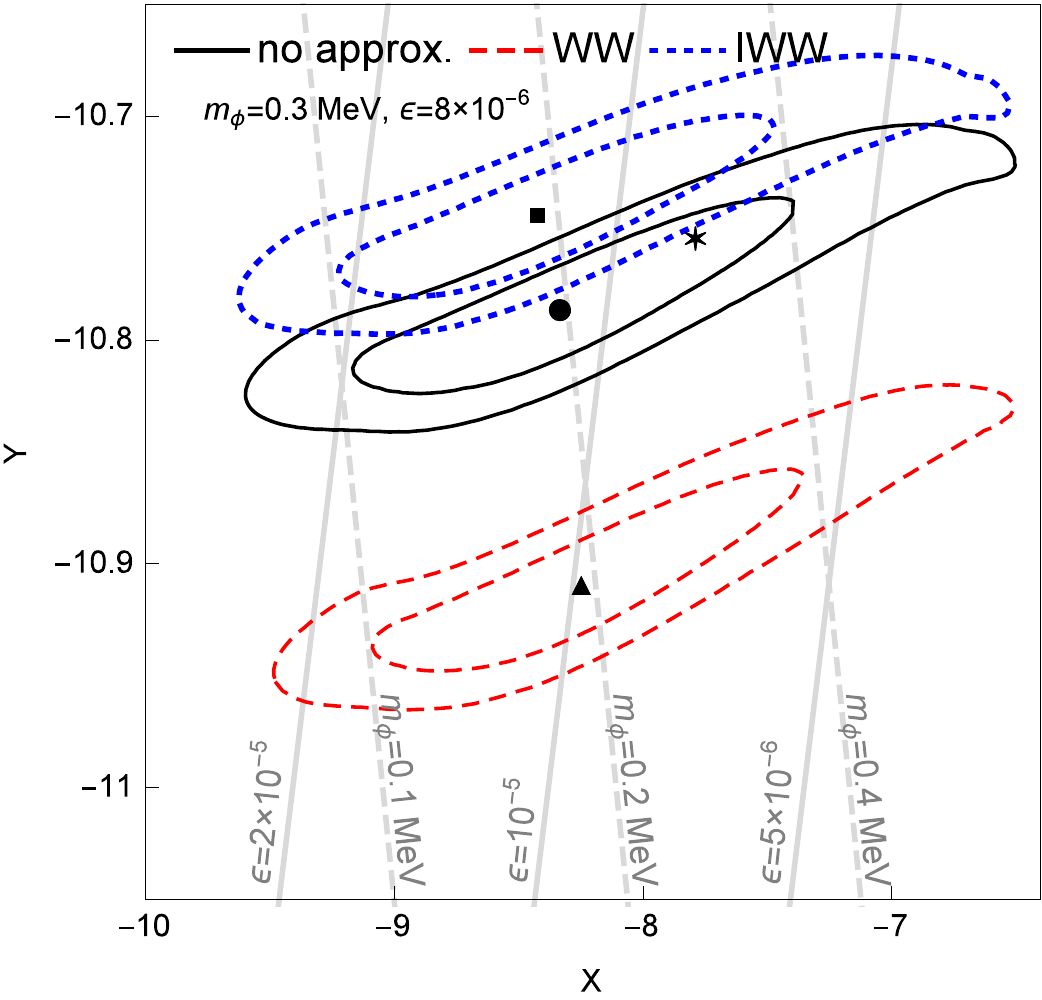}}
\caption{\label{fig:pseudo_exp_3} Assuming the scalar boson exists with $(m_\phi,\epsilon)=(0.3{\rm\;MeV,8\times 10^{-6}})$ and is observed in E137 with 137 times luminosity. (a)--(c) See the caption in Fig. \ref{fig:pseudo_exp_1} for details. (d) Change of coordinate of $\chi^2$ fit plot: $X=\ln\frac{m_0}{\rm 1\;GeV}+\ln\frac{m_\phi}{m_0}\cos\theta-\ln\frac{\epsilon}{\epsilon_0}\sin\theta$ and $Y=\ln\epsilon_0+\ln\frac{m_\phi}{m_0}\sin\theta+\ln\frac{\epsilon}{\epsilon_0}\cos\theta$, where $\theta=42.4^{\circ}$, $m_0=0.1$ MeV, and $\epsilon_0=2\times10^{-5}$. This means to rotate the coordinate $42.4^{\circ}$ with respect to $(m_\phi,\epsilon)=(0.1{\rm\;MeV,2\times 10^{-5}})$.}
\end{figure}

\begin{figure}
\centering
\includegraphics[scale=0.8]{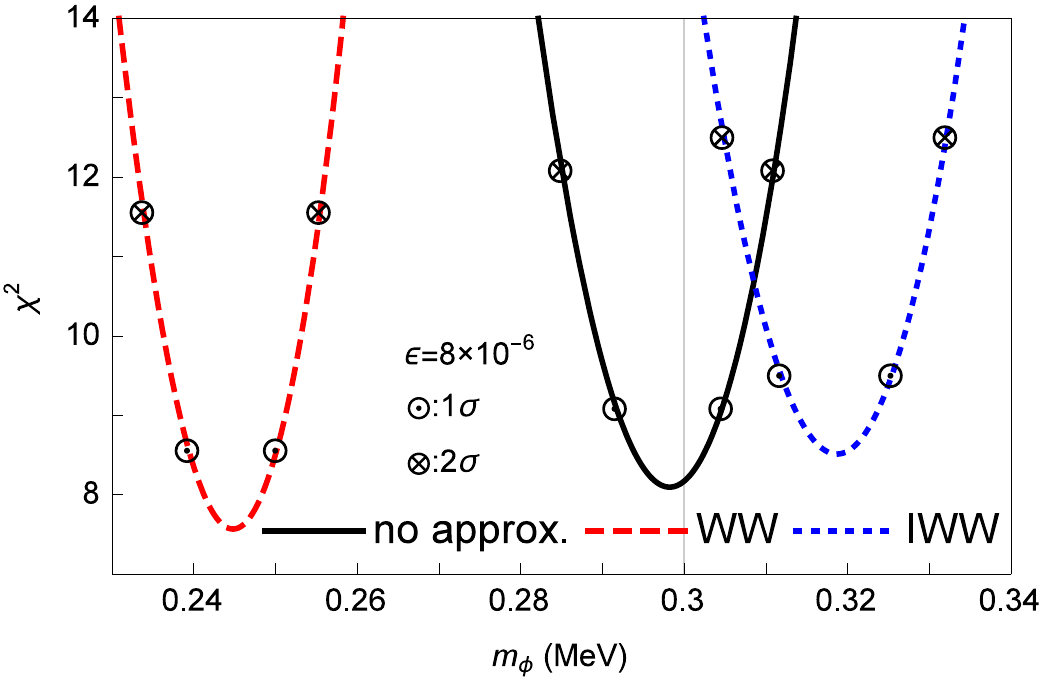}
\caption{\label{fig:pseudo_exp_3_2} Assuming the scalar boson exists with $(m_\phi,\epsilon)=(0.3{\rm\;MeV,8\times 10^{-6}})$ and is observed in E137 with 137 times luminosity. The number of events distribution is the same in Fig. \ref{fig:pseudo_exp_3}. The value of $\chi^2$ with respect to $m_\phi$ (assuming $\epsilon$ is precisely measured): the black, dashed red, and dotted blue lines correspond to the $\chi^2$ values calculated with no, WW, and IWW approximation. The minimum of $\chi^2$ corresponds the best fit $m_\phi$; the circle dots $\odot$ correspond to 1$\sigma$ range ($\Delta\chi^2=1$); the circle crosses $\otimes$ correspond to 2$\sigma$ range ($\Delta\chi^2=4$). The gray vertical line indicates the true value of $m_\phi$.}
\end{figure}

We show the results of these pseudoexperiments with $(m_\phi,\epsilon)=(110{\rm\;MeV,10^{-7}})$ in Fig. \ref{fig:pseudo_exp_1}, $(m_\phi,\epsilon)=(200{\rm\;MeV,1.3\times10^{-7}})$ in Fig. \ref{fig:pseudo_exp_2}, and $(m_\phi,\epsilon)=(0.3{\rm\;MeV,8\times 10^{-6}})$ in Fig. \ref{fig:pseudo_exp_3}. We see that the ``true'' parameter values lie within the $1\sigma$ allowed regions when fitting with the complete calculation. On the other hand, although using approximation sometimes gives a fairly good estimate of cross section, the result of data fitting lies outside the $2\sigma$ range. It is worth noting that the shape of the $1\sigma$ or $2\sigma$ range is roughly along the exclusion boundary in Fig. \ref{fig:E137}, because the exclusion boundary is the isocontour of the number of events.

Next, we consider another scenario of the third pseudoexperiment with $(m_\phi,\epsilon)=(0.3{\rm\;MeV,8\times 10^{-6}})$. In this part of parameter space, the allowed coupling and mass can extend over roughly an order of magnitude. To illustrate the usefulness of the complete calculation, we perform fits to this data assuming that there is another experimental result that can sensitively measure the coupling. This would be the case if recently proposed experiments involving decays of radioactive nuclei underground see a nonzero signal~\cite{Izaguirre:2014cza,Liu:2016qwd} and we can use the beam dump experiment to determine the mass precisely. For simplicity, we assume that the other experiment measures the coupling with negligible error. Since there is one parameter to fit, the $1\sigma$ and $2\sigma$ range correspond to $\Delta\chi^2=1$ and $\Delta\chi^2=4$. We show the results in Fig. \ref{fig:pseudo_exp_3_2}. Again as expected, we see that the ``true'' parameter values lie within the $1\sigma$ allowed region when fitting with the complete calculation. Using the approximations, the ``true'' mass lies outside the $2\sigma$ ranges. We observe that using the complete calculation could be crucial in measuring the mass of a new particle in this region of parameter space.

\section{discussion}\label{sec:discussion}

Our results are based on a new scalar boson motivated by proton radius puzzle~\cite{Liu:2016qwd}. However, we expect that the qualitative description remains similar in other type of particles, such as pseudoscalar and vector. While the production amplitude, decay length, and the absorption length can differ in detail for particles with different quantum numbers, they are qualitatively similar. The approximations that we have examined deal with the phase space integral and coupling to electromagnetism of the target nucleus. Therefore, we expect similar results to hold for other bosons as in the scalar case. The similarity of our exclusion plot to the vector case \cite{Bjorken:2009mm,Andreas:2012mt} provides evidence in favor of this. Including a coupling to the muon may change the situation for $m_\phi>2m_\mu$~\cite{Liu:2016qwd} due to the opening of a new channel with typically a substantial partial width. A study of the production of vector particles in electron beam dumps that deals with some of the issues we have addressed can be found in Ref.~\cite{Beranek:2013yqa}.

There are some other beam dump experiments using a Cherenkov detector, such as E141 \cite{Riordan:1987aw} and Orsay \cite{Davier:1989wz}. Their exclusion plots do not extend to the region where $m_\phi<2m_e$. We show the results of the beam dump experiments E141 and Orsay in Ref.~\cite{Liu:2016qwd}.

In this work, we present a complete analysis of beam dump experiments. We show that a brute-force analytical calculation is possible. Software exists using Monte-Carlo simulations, such as \textsc{MadGraph/MadEvent}~\cite{Alwall:2007st} as used in, e.g.,~\cite{Essig:2010xa}, that can calculate the cross section without using approximations. Our work can be used as a consistency check for Monte-Carlo simulations. We show that using the WW approximation can be trusted to an order of magnitude in cross sections and exclusion plots. Additionally our work  allows us to understand the errors introduced by the various common approximations. In certain regions of parameter space different errors partially cancel against each other, leading to results that are accidentally sometimes more accurate than might be expected. However, as we illustrated with several pseudoexperiments in a range of masses, in the event of a nonzero signal, a complete calculation can give very different results from the approximations. This could be useful given the possibility of future electron beam dump experiments~\cite{future}.

\section*{Acknowledgement}
We acknowledge J. Detwiler and R. Essig for invaluable discussions and suggestions. The work of G. A. M. and Y.-S. L. was supported by the U. S. Department of Energy Office of Science, Office of Nuclear Physics under Award Number DE-FG02-97ER-41014. The work of D.M. was supported by the U.S. Department of Energy under Grant No. DE-SC0011637.


\begin{thebibliography}{00}
  
\bibitem{Essig:2013lka} 
  R.~Essig, J.~A.~Jaros, W.~Wester, P.~H.~Adrian, S.~Andreas, T.~Averett, O.~Baker, B.~Batell {\it et al.},
  Working group report: New light weakly coupled particles,
  arXiv:1311.0029.


\bibitem{Bjorken:2009mm} 
  J.~D.~Bjorken, R.~Essig, P.~Schuster, and N.~Toro,
  New fixed-target experiments to search for dark gauge forces,
  Phys.\ Rev.\ D {\bf 80}, 075018 (2009).

\bibitem{Andreas:2012mt} 
  S.~Andreas, C.~Niebuhr, and A.~Ringwald,
  New limits on hidden photons from past electron beam dumps,
  Phys.\ Rev.\ D {\bf 86}, 095019 (2012).

\bibitem{Blumlein:2013cua} 
  J.~Blümlein and J.~Brunner,
  New exclusion limits on dark gauge forces from proton bremsstrahlung in beam-dump data,
  Phys.\ Lett.\ B {\bf 731}, 320 (2014).


\bibitem{deNiverville:2016rqh} 
  P.~deNiverville, C.~Y.~Chen, M.~Pospelov, and A.~Ritz,
  Light dark matter in neutrino beams: production modelling and scattering signatures at MiniBooNE, T2K and SHiP,
  arXiv:1609.01770.


\bibitem{vonWeizsacker:1934nji} 
  C.~F.~von Weizs\"{a}cker,
  Radiation emitted in collisions of very fast electrons,
  Z.\ Phys.\  {\bf 88}, 612 (1934).
  
\bibitem{Williams:1935dka} 
  E.~J.~Williams,
  Correlation of certain collision problems with radiation theory,
  Kong.\ Dan.\ Vid.\ Sel.\ Mat.\ Fys.\ Med.\  {\bf 13N4}, no. 4, 1 (1935).
 

\bibitem{Kim:1973he} 
  K.~J.~Kim and Y.~S.~Tsai,
  Improved Weizs\"{a}cker-williams method and its application to lepton and W boson pair production,
  Phys.\ Rev.\ D {\bf 8}, 3109 (1973).

\bibitem{Tsai:1973py} 
  Y.~S.~Tsai,
  Pair production and bremsstrahlung of charged leptons,
  Rev.\ Mod.\ Phys.\  {\bf 46}, 815 (1974); {\bf 49}, 421(E) (1977).


\bibitem{Tsai:1986tx} 
  Y.~S.~Tsai,
  Axion bremsstrahlung by an electron beam,
  Phys.\ Rev.\ D {\bf 34}, 1326 (1986).


\bibitem{Brodsky:1971ud}
  S.~J.~Brodsky, T.~Kinoshita, and H.~Terazawa,
  Two photon mechanism of particle production by Hhgh-energy colliding beams,
  Phys.\ Rev.\ D {\bf 4}, 1532 (1971).
  
  
\bibitem{Bjorken:1988as} 
  J.~D.~Bjorken, S.~Ecklund, W.~R.~Nelson, A.~Abashian, C.~Church, B.~Lu, L.~W.~Mo, T.~A.~Nunamaker, and P. Rassmann,
  Search for neutral metastable penetrating particles produced in the SLAC beam dump,
  Phys.\ Rev.\ D {\bf 38}, 3375 (1988).


\bibitem{TuckerSmith:2010ra} 
  D.~Tucker-Smith and I.~Yavin,
  Muonic hydrogen and MeV forces,
  Phys.\ Rev.\ D {\bf 83}, 101702 (2011).
   
\bibitem{Liu:2016qwd} 
  Y.~S.~Liu, D.~McKeen, and G.~A.~Miller,
  Electrophobic scalar boson and muonic puzzles,
  Phys.\ Rev.\ Lett.\  {\bf 117}, 101801 (2016).


\bibitem{DeJager:1987qc} 
  H.~De Vries, C.~W.~De Jager, and C.~De Vries,
  Nuclear charge and magnetization density distribution parameters from elastic electron scattering,
  Atom.\ Data Nucl.\ Data Tabl.\  {\bf 36}, 495 (1987).


\bibitem{atomic form factor}
P. J. Brown, A. G. Fox, E. N. Maslen, M. A. O'Keefe, and B. T. M. Willis, \textit{International Tables for Crystallography} (2006), Vol. C, ch. 6.1, pp. 554-595.


\bibitem{Tsai:1966js}
  Y.~S.~Tsai and V.~Whitis,
  Thick target bremsstrahlung and target consideration for secondary particle production by electrons,
  Phys.\ Rev.\  {\bf 149}, 1248 (1966).


\bibitem{Pospelov:2008zw} 
  M.~Pospelov,
  Secluded U(1) below the weak scale,
  Phys.\ Rev.\ D {\bf 80}, 095002 (2009).

\bibitem{Bouchendira:2010es} 
  R.~Bouchendira, P.~Clade, S.~Guellati-Khelifa, F.~Nez, and F.~Biraben,
  New determination of the fine structure constant and test of the quantum electrodynamics,
  Phys.\ Rev.\ Lett.\  {\bf 106}, 080801 (2011).
  
  
\bibitem{Eides:2000xc} 
  M.~I.~Eides, H.~Grotch, and V.~A.~Shelyuto,
  Theory of light hydrogen-like atoms,
  Phys.\ Rept.\  {\bf 342}, 63 (2001).


\bibitem{Izaguirre:2014cza} 
  E.~Izaguirre, G.~Krnjaic, and M.~Pospelov,
  Probing new physics with underground accelerators and radioactive sources,
  Phys.\ Lett.\ B {\bf 740}, 61 (2015).


%
\bibitem{Beranek:2013yqa} 
  T.~Beranek, H.~Merkel, and M.~Vanderhaeghen,
  Theoretical framework to analyze searches for hidden light gauge bosons in electron scatteri7g fixed target experiments,
  Phys.\ Rev.\ D {\bf 88}, 015032 (2013).
  

\bibitem{Riordan:1987aw} 
  E.~M.~Riordan {\it et al.},
  A search for short lived axions in an electron beam dump experiment,
  Phys.\ Rev.\ Lett.\  {\bf 59}, 755 (1987).


\bibitem{Davier:1989wz} 
  M.~Davier and H.~Nguyen Ngoc,
  An unambiguous search for a light Higgs boson,
  Phys.\ Lett.\ B {\bf 229}, 150 (1989).


\bibitem{Alwall:2007st} 
  J.~Alwall, P.~Demin, S.~de Visscher, R.~Frederix, M.~Herquet, F.~Maltoni, T.~Plehn, D.~L.~Rainwater, and T.~Stelzer,
  MadGraph/MadEvent v4: The new web generation,
  J. High Energy Phys. 09 ({\bf 2007}) 028.


%
\bibitem{Essig:2010xa} 
  R.~Essig, P.~Schuster, N.~Toro, and B.~Wojtsekhowski,
  An electron fixed target experiment to search for a new vector boson A' decaying to e+e-,
  J. High Energy Phys. 02 ({\bf 2011}) 009.
  

  
\bibitem{future}
  B.~Wojtsekhowski,
  Searching for a U-boson with a positron beam,
  AIP Conf.\ Proc.\  {\bf 1160}, 149 (2009);
  S.~Abrahamyan {\it et al.} (APEX Collaboration),
  Search for a new gauge boson in electron-nucleus fixed-target scattering by the APEX experiment,
  Phys.\ Rev.\ Lett.\  {\bf 107}, 191804 (2011);
  E.~Izaguirre, G.~Krnjaic, P.~Schuster, and N.~Toro,
  New electron beam-dump experiments to Search for MeV to few-GeV dark matter,
  Phys.\ Rev.\ D {\bf 88}, 114015 (2013);
  M.~Raggi and V.~Kozhuharov,
  Proposal to search for a dark photon in positron on target collisions at DA$\Phi$NE linac,
  Adv.\ High Energy Phys.\  {\bf 2014}, 959802 (2014);
  E.~Izaguirre, G.~Krnjaic, P.~Schuster, and N.~Toro,
  Testing GeV-scale dark matter with fixed-target missing momentum experiments,
  Phys.\ Rev.\ D {\bf 91}, 094026 (2015).


\end{thebibliography}
\end{document}